\documentclass[prb, 10pt, twocolumn, aps, showpacs,preprintnumbers,amsmath,amssymb, longbibliography]{revtex4-1}

\usepackage{graphicx}
\usepackage{dcolumn}
\usepackage{bm}
\usepackage[bookmarks, colorlinks=true, breaklinks]{hyperref}
\hypersetup{linkcolor=blue,citecolor=blue,filecolor=black,urlcolor=blue}
\usepackage{color} 

\newcommand{\scf}{ScF$_3$}
\newcommand{\reo}{ReO$_3$}
\newcommand{\wo}{WO$_3$}
\newcommand{\alf}{AlF$_3$}
\newcommand{\tif}{TiF$_3$}

\makeatletter
\renewcommand*\env@matrix[1][*\c@MaxMatrixCols c]{%
  \hskip -\arraycolsep
  \let\@ifnextchar\new@ifnextchar
  \array{#1}}
\makeatother

\emergencystretch=\maxdimen
\begin{document}

\title{Landau theory and  giant room-temperature barocaloric effect in MF$_3$ metal trifluorides} 

\author{A. Corrales-Salazar$^1$, R. T. Brierley$^2$, P. B. Littlewood$^{3,4}$,  G. G. Guzm\'{a}n-Verri$^{1,4}$}
\affiliation{$^{1}$Materials Research Science and Engineering Center, University of Costa Rica, San Jos\'{e}, Costa Rica 11501,}
\affiliation{$^{2}$Department of Physics, Yale University, New Haven, Connecticut, USA, 06511,}
\affiliation{$^{3}$James Franck Institute, University of Chicago, 929 E 57 St, Chicago, Illinois, USA 60637,}
\affiliation{$^{4}$Materials Science Division, Argonne National Laboratory, Argonne, Illinois, USA 60439,}

\date{\today}

\begin{abstract}
The structural phase transitions of MF$_3$~(M=Al, Cr, V, Fe, Ti, Sc) metal trifluorides
are studied within a simple Landau theory consisting of tilts of rigid MF$_6$ octahedra
associated with soft antiferrodistoritive optic modes that are coupled to 
long-wavelength strain generating acoustic phonons.
We calculate the temperature and pressure dependence of several quantities such as 
the spontaneous distortions, volume expansion and shear strains as well as 
$T-P$ phase diagrams. 
By contrasting our model to experiments we quantify the deviations from mean-field behavior 
and found that the tilt fluctuations of the MF$_6$ octahedra
increase with metal cation size.
We apply our model to predict giant barocaloric effects in Sc substituted \tif\
of up to about $15\,$JK$^{-1}$kg$^{-1}$  for modest
hydrostatic compressions of  $0.2\,$GPa.
The effect extends over a wide temperature range of over $140\,$K (including room temperature) 
due to a large predicted rate $dT_c/dP = 723\,$K GPa$^{-1}$, which exceeds those of typical barocaloric materials.
Our results suggest that open lattice frameworks such as the trifluorides are an attractive 
platform to search for giant barocaloric effects. 
\end{abstract}

\maketitle

\section{Introduction}

Metal trifluorides (or simply trifluorides)  are a class of materials 
with chemical formula MF$_3$~(M=Al, Cr, V, Fe, Ti, Sc)  and with an open lattice 
framework in which the trivalent metal ion M is surrounded by an octahedron of corner-shared fluorine atoms.~\cite{Hepworth1957a, Daniel1990a}  
They are isostructural to \reo\, a well-known ABO$_3$ perovskite in which the A site is vacant.~\cite{Tsuda2000a}
They can exhibit large thermal expansion~(TE)  which 
can be reversibly tuned from positive~(PTE) to negative~(NTE) by 
temperature, pressure, cation substitution,  or
redox intercalation.~\cite{Morelock2013a, Morelock2014a, Hu2014a, Morelock2015a, Romao2015a, Wang2016a, Chen2017a} 
This makes the trifluorides attractive for designing materials that 
are dimensionally stable and resistant to thermal shocks.~\cite{Wang2015a, *Chen2015a, *Romao2013b, *Lind2012a, *Miller2009a} 

 At ambient pressure, most trifluorides exhibit antiferrodistortive structural transitions 
 with cubic-to-rhombohedral~($c-r$) transformations in which
 the MF$_6$ octahedron tilts around the $(111)$ axis. The tilting angles are large, e.g., about 14$^\circ$ in \alf\
at room temperature~(RT)~\cite{Morelock2014b} and are accompanied by 
spontaneous shear and volume strains.~\cite{Kennedy2002a} 
Such lattice instability is the result from the
condensation of a three-fold zone-boundary $R_4^+$ phonon mode of 
the cubic phase located at the wavevector $(1,1,1)(\pi/a)$.~\cite{Chen2004a}
Below the transition, the  $R_4^+$ mode
splits into a low energy E$_g$ doublet and a high energy A$_{1g}$ singlet.~\cite{Daniel1990b}  

Density functional theory,~\cite{Chen2004a} molecular dynamics~(MD) simulations,~\cite{Chaudhuri2004a} 
and electrostatic energy considerations~\cite{Allen2006a} 
have shown that the driving force of the lattice instability in the trifluoride is of dipolar origin.
When the M-F-M bond bends, fluorine displaces transverse to the bond length
generating an electric dipole with 
a negative end at the F$^{-}$ anion and a positive end at 
its cubic lattice site. 
\begin{figure}[htp!]
	\includegraphics[scale=0.24]{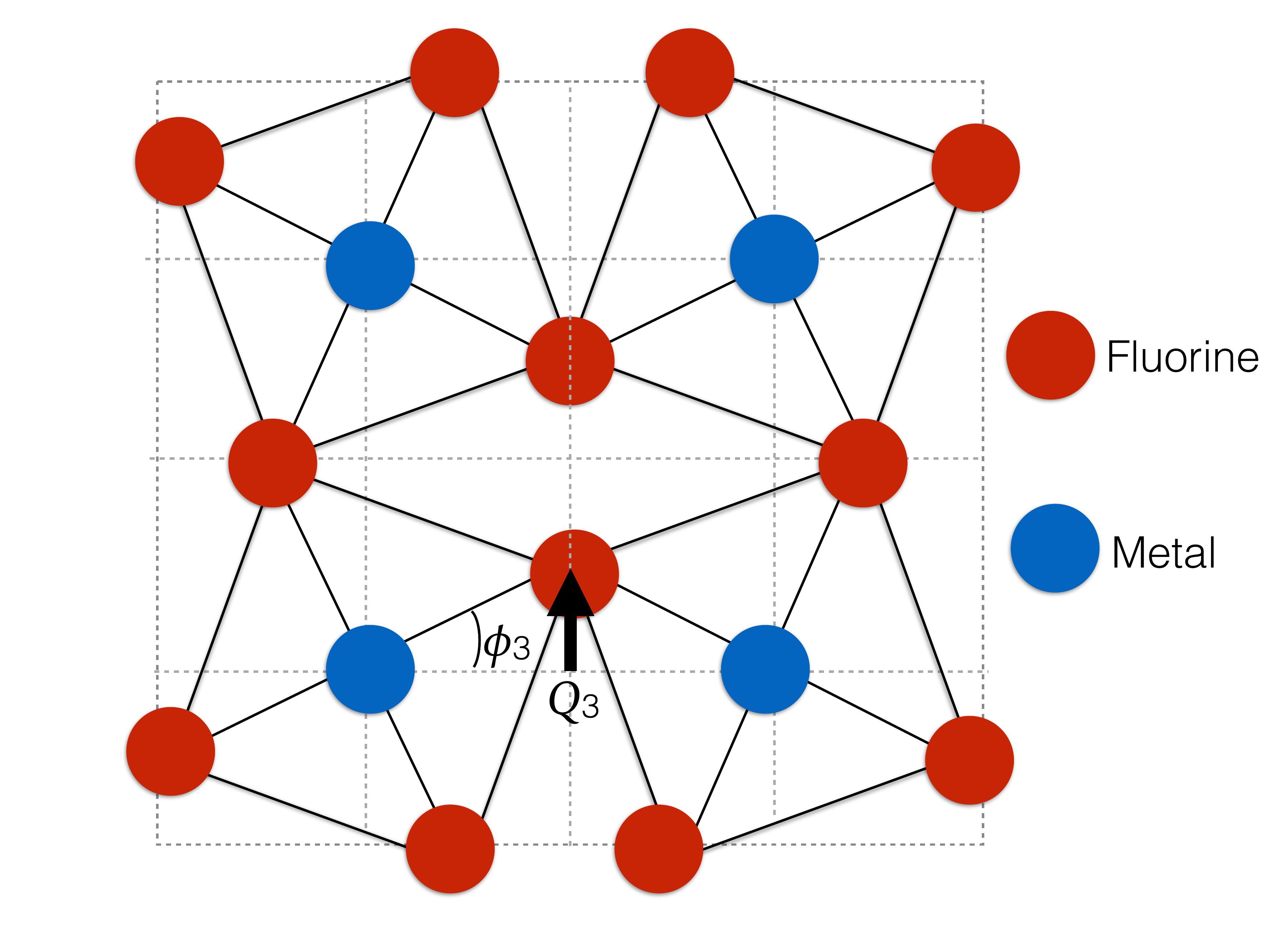}
	\caption{A $(001)$ section of the MF$_3$ lattice illustrating
	the displacement of fluorine ions described by the soft-mode coordinate $Q_3$ 
	and the antiferrodistortive rotation $\phi_3$ of the MF$_6$ octahedra around $(001)$.  }
	\label{fig:phi}
\end{figure}
This distortion concomitantly induces a polarization in the fluorine electron cloud
that is opposite to the displacive dipole.  
While there is an energy penalty for creating such induced dipoles, 
the resulting interactions between the induced-dipoles and between the induced-dipoles with the ionic charges
lower the total energy to favor the r-phase over the parent c-structure preferred by the 
purely ionic  Madelung energy. 
 
A trifluoride of special recent interest is \scf ,
an ionic  insulator with a wide indirect energy band gap of about  $8-10\,$eV.~\cite{Bocharov2016a, *Hamed2015a}
It does not have a structural transition to an $r$ phase at ambient pressure but rather
exhibits incipient behavior 
in which a nearly flat M-R phonon branch softens
without condensing, as it has been observed by inelastic x-ray 
scattering experiments~(IXS)~\cite{Handunkanda2015a}
and found in ab-initio calculations.~\cite{Roekeghem2016a}
It exhibits strong negative TE (-34 ppm K$^{-1}$ near RT)
from $10-1100\,$K~\cite{Greve2010a}  and 
and very strong lattice anharmonicities
~(its soft $R_4^+$ mode is described by a quartic potential energy in the tilts).~\cite{Li2011a}
Its incipient behavior and proximity to a r-phase induced
by, e.g., cation substitution~\cite{Morelock2013a, Morelock2014a, Morelock2015a}
suggest that \scf\ is one of the few known stoichiometric materials
near a quantum structural phase transition.~\cite{Handunkanda2015a} 

With hydrostatic compression, the $r$ phase can be induced at
higher temperatures.
For example, at about $0.6\,$GPa,  a $c-r$ transition is observed  near RT in the incipient  \scf .~\cite{Greve2010a, Aleksandrov2009a, Aleksandrov2002a}
X-ray diffraction experiments have determined the temperature-pressure phase diagrams 
for Sc substituted \alf ~(Sc$_{1-x}$Al$_x$F$_3$).~\cite{Morelock2015a}  
Very significantly, they have observed linearly 
increasing transition temperatures with pressure
with large rates ($dT_c / d P \simeq 400-500\,$K GPa$^{-1}$)
that vary little with Sc concentration and pressure.~\cite{Greve2010a, Morelock2015a}
Additional pressure-induced transitions have been reported at higher pressures.~\cite{Aleksandrov2009a, Aleksandrov2002a}  

While microscopic models for the trifluoride are available,~\cite{Chen2004a, Chaudhuri2004a, Allen2006a}
 there is currently no macroscopic approach based on the simple Landau phenomenology. 
The purpose of this work is thus to construct such a theory.
Our model consists of rigid tilts of the MF$_6$ octahedra 
associated with the soft $R_4^+$ optic mode coupled to 
long-wavelength strain generating acoustic phonons.
The model is similar to those used to describe the widely studied antiferrodistortive 
transitions of SrTiO$_3$ and LaAlO$_3$,~\cite{Cowley1980a, Slonczewski1970a} but 
with the important 
distinctions that in the trifluorides the phase transition can be discontinuous and 
that there are large excess volume strains.
By comparing our model to experiments on several trifluorides 
we quantify the deviations from mean-field behavior 
and found trends with the metal cation size. 

We also apply our model to predict the barocaloric effect (BCE) in the trifluorides. 
BCEs are reversible thermal changes in a substance in response 
to changes in hydrostatic pressure
and are currently of enormous interest for their potential in developing clean and efficient
solid-state cooling technologies.~\cite{Manosa2017a, *Lu2015a, *Moya2014a}  
It is expected that materials with strong TE such as the trifluorides 
should give rise to large barocaloric responses, as their
entropy rate $(\partial S / \partial P)_T = - (\partial V / \partial T)_P$,
according to the Maxwell's relations.~\cite{Moya2014a}
Indeed, we show that the isothermal changes entropy in 
the trifluorides are comparable to those of other classes of materials 
exhibiting so-called giant BCEs,~\cite{Manosa2010a, Manosa2011a, Flerov2011a, Suheyla2012a, Stern2014a, Matsunami2015a, Lloveras2015a, Stern-Taulats2016a, BermudezGarcia2017a} 
and that it can extend over a broad temperature range which includes RT for modest changes in pressure
as a result of their large barocaloric coefficients $dT_c/dP$. 
So far, the BCE in the trifluoride has not been studied neither experimentally nor theoretically.

This paper is organized as follows: In Sec. II we present our Landau theory to describe the structural 
transitions; in Sec. III
we show our results and discussion including a comparison to the isostructural compounds \reo\ and \wo  ;
and in Sec. IV we present our conclusions.

\section{Landau Theory}

\subsection{Free energy}

We choose  the order parameter as the linear 
displacement ${\bm Q}=(Q_1,Q_2,Q_3)$ which represents, in first order,
an antiferrodistortive rotation of the MF$_6$ octahedra  through angles  $ \phi_i $ and $ -\phi_i, ~(i=1,2,3)$ 
about axes parallel to a cube edge. We normalize the $Q$'s in such a way that they are numerically
equal to the linear fluorine displacements. They are related to $\phi_i$ 
by $ \tan \phi_i =  2Q_i/a$, where $a$
is the lattice constant, see Fig.~\ref{fig:phi}. 
In addition to the antiferrodistortive distortion,
we introduce elastic strains $\eta_{\alpha}$  as a secondary order parameter.  
We  write the components of the strain tensor  in the usual Voigt notation:
$ \eta_\alpha  \equiv  \epsilon_{\alpha \alpha} = \partial u_\alpha / \partial x_\alpha~(\alpha=1,2,3)$,  $ \eta_4 = 2\epsilon_{yz} = 2 \left( \partial u_y / \partial z +  \partial u_z / \partial y \right),  \eta_5 = 2\epsilon_{xz} = 2 \left( \partial u_x / \partial z +  \partial u_z / \partial x \right)$, and  $\eta_6 = 2\epsilon_{xy} = 2 \left( \partial u_x / \partial y +  \partial u_y / \partial x \right)$.
We do not consider fluctuations in ${\bm Q}$ and $\eta_{\alpha}$.

Our Landau free energy density is given as follows:
\begin{align}
	\label{eq:free_energy}
	 G_Q+G_{\eta} + P \sum_{\alpha=1}^{3} \eta_{\alpha},
\end{align}
where $G_Q$ is a strain-free free energy,
\begin{align}
	\label{eq:gibs}
	 G_Q &= G_0 + \frac{A}{2} \left(Q_1^2 + Q_2^2 + Q_3^2\right) \nonumber \\ 
	 	& + \frac{u}{2} \left(Q_1^2 + Q_2^2 + Q_3^2\right)^2 + \frac{3v}{2}  \left( Q_1^2 Q_2^2 + Q_1^2 Q_3^2 +Q_2^2 Q_3^2 \right)  \\ 
         &+ \frac{w_1}{6}  \left( Q_1^2 + Q_2^2 + Q_3^2 \right)^3,  \nonumber 
\end{align}
where $A=A_0(T-T_0)$ and $T_0$ is the supercooling temperature that limits of stability of the parent  c-phase.
$G_{\eta}$ is an energy density with elastic couplings,
\begin{align}
 \label{eq:Ge}
 G_{\eta} &=   e_a ( \eta_1 + \eta_2 + \eta_3)\left(Q_1^2 + Q_2^2 + Q_3^2\right) \\  \nonumber 
       &- e_t \left[ \eta_1 \left( 2 Q_1^2-Q_2^2-Q_3^2 \right) + \eta_2 \left( 2 Q_2^2-Q_1^2-Q_3^2  \right) \right. \\ \nonumber 
       &~~~~~~~~~~~~~~~~~~~~~~~~~~~~~~~~~~~~~
       \left. + \eta_3 \left( 2 Q_3^2-Q_1^2-Q_2^2  \right) \right] \\ \nonumber
       & - e_r ( Q_1 Q_2 \eta_6 + Q_1 Q_3 \eta_5 + Q_2 Q_3 \eta_4) + \frac{1}{2} \sum_{\alpha \beta} C_{\alpha \beta}^0 \eta_{\alpha} \eta_{\beta}.
\end{align}
$A_0,u,v,w, e_a,e_t$, and $e_r$ are model parameters independent of temperature and pressure 
and $C_{\alpha \beta}^0$ are the usual elastic constants of the parent phase in the Voigt notation. 
The third term in the free energy (\ref{eq:free_energy}) is a hydrostatic compression where $P$
is measured from atmospheric pressure. 

In writing the free energy~(\ref{eq:free_energy}), we have not considered
any polar degrees of freedom associated with phonon modes that break inversion symmetry 
as there is no evidence that such lattice modes are unstable in the trifluorides, e.g., the zone-center 
TO phonon modes do not condense and remain fairly energetic such as in \scf\ ($4-5\,$THz)
~\cite{Roekeghem2016a, Piskunov2016a, Liu2015a} and other trifluorides.~\cite{Zinenko2000a}
Moreover, Clausius-Mossotti theory predicts that the ground state 
exhibits antipolar order from the MF$_6$ tilts with null polarization.~\cite{Allen2006a}
We have also ignored sixth-order cubic anisotropies.  
Our results will show that this is justified as long as we are describing the  r-phase. 
In Appendix B, we show, however, that they are essential to describe other pressure-induced phases.
We have also neglected any polar degrees of freedom associated with phonon modes 
that would break inversion symmetry, as there is no evidence that such lattice modes are unstable in the trifluorides.
 
Minimizing Eq.~(\ref{eq:free_energy}) with respect to the strains gives,
\begin{align} \label{eq:etas}
  \eta_1 &= - \frac{e_a}{3C_a}\left(Q_1^2+Q_2^2+Q_3^2\right) \nonumber \\ 
        &~~~~~~~~~~~~~~~~~~
        +\frac{e_t}{2C_t}(2Q_1^2-Q_2^2-Q_3^2) - \frac{P}{3C_a}, \nonumber  \\
  \eta_2 &= -\frac{e_a}{3C_a}\left(Q_1^2+Q_2^2+Q_3^2\right) \nonumber \\ 
        &~~~~~~~~~~~~~~~~~~
        +\frac{e_t}{2C_t}(2Q_2^2-Q_1^2-Q_3^2) - \frac{P}{3C_a}, \nonumber  \\
  \eta_3 &= - \frac{e_a}{3C_a}\left(Q_1^2+Q_2^2+Q_3^2\right) \nonumber \\ 
        &~~~~~~~~~~~~~~~~~~
        +\frac{e_t}{2C_t}(2Q_3^2-Q_1^2-Q_2^2) - \frac{P}{3C_a}, \\
 \eta_4 &=  \frac{e_r}{C_r} Q_2 Q_3, \nonumber  \\
 \eta_5 &=  \frac{e_r}{C_r} Q_1 Q_3, \nonumber \\
 \eta_6 &=  \frac{e_r}{C_r} Q_1 Q_2, \nonumber 
 \end{align}
 where $C_a=  \left(1/3 \right) \left( C_{11}^0+2C_{12}^0 \right)$ is 
 the bulk modulus, $C_t = \left(1/2 \right) \left( C_{11}^0-C_{12}^0 \right)$, 
 and $C_r= C_{44}^0$ are the shear tetragonal and rhombohedral moduli, respectively.

When the spontaneous strains of Eq.~(\ref{eq:etas}) are substituted back into Eq.~(\ref{eq:free_energy}), 
we obtain,  as expected,~\cite{Cowley1980a} that 
the free energy has the same form as that of Eq.~(\ref{eq:gibs}) 
for the strain-free case except with renormalized quadratic ($A$) and quartic coefficients ($u$ and $v$)
and a uniform energy shift due to pressure,
\begin{align}
\label{eq:gibsRenormalized}
\tilde{G}(T,P) &= G_0 + \frac{1}{2} \tilde{A}\left(Q_1^2 + Q_2^2 + Q_3^2\right) \nonumber \\ 
   & + \frac{\tilde{u}}{2} \left(Q_1^2 + Q_2^2 + Q_3^2\right)^2 \nonumber \\ 
   &+  \frac{3\tilde{v}}{2} \left( Q_1^2 Q_2^2 + Q_1^2 Q_3^2 + Q_2^2 Q_3^2 \right)   \\ 
   &+ \frac{w_1}{6} \left( Q_1^2 + Q_2^2 + Q_3^2 \right)^3  \nonumber \\ 
    & -\frac{1}{2} \frac{P^2}{C_a}, \nonumber
\end{align}
where,
\begin{align}
\label{eq:ARenormalized}
 \tilde{A} = A - \frac{2 e_a P}{C_a},
\end{align}
and, 
\begin{subequations}
 \label{eq:uandv}
\begin{align}
 \tilde{u} = u -  \left( 5  \frac{e_a^2}{  C_a } + 3  \frac{e_t^2}{ C_t} \right), \\
 \tilde{v} = v + \left( 3  \frac{ e_t^2 }{ C_t } - \frac{1}{3} \frac{e_r^2}{ C_r } \right). 
\end{align}
\end{subequations}
We conclude the presentation of the free energy here. 
In the next section we  apply our model to describe the $c-r$ transition
of the trifluorides.

\subsection{$c-r$ transition}

The symmetry of the ground state and order of the phase transition is determined by
the choice of $ \tilde{u} $ and  $\tilde{v}$. For a $c-r$ discontinuous~(continuous) transition, we must have
$ \tilde{v} <0$ and $\tilde{u} +  \tilde{v}<0$~($\tilde{u} +  \tilde{v}>0$).~\cite{Cowley1980a}

To describe the r-phase, we take ${\bm Q}=(Q_s / \sqrt{3})(1,1,1)$, where $Q_s$ is determined by 
minimization of the free energy (\ref{eq:gibsRenormalized}),
\begin{align}
\label{eq:Qs}
 Q_s(T,P)  & = \pm  \left\{ \sqrt{    \left( \frac{  \tilde{u} +  \tilde{v} }{ w_1 } \right)^2 - \frac{\tilde{A}}{w_1 }  }  - \left( \frac{  \tilde{u} +  \tilde{v} }{ w_1 } \right)  \right\}^{1/2}.
\end{align}
Substitution of the order parameter (\ref{eq:Qs}) into Eq.~(\ref{eq:etas}) gives, respectively, the following
the spontaneous volume and shear strains,
\begin{subequations}
\label{eq:etas1}
\begin{align}
\eta_a &= \eta_1 + \eta_2 + \eta_3 =  -\frac{e_a}{C_a} Q_s^2 - P/C_a, \\
\label{eq:etas1b}
\eta_r &= \eta_4 = \eta_5 = \eta_6 =  \frac{e_r}{3C_r} Q_s^2 = \cos \alpha_C,
\end{align}
\end{subequations}
where $\alpha_C$ as the angle between any two axes of the $c$ unit cell. 

Experiments~\cite{Morelock2015a, Morelock2014a, Morelock2014b} usually 
report the ratio between the lattice constants $c_H$  and $a_H$ of a hexagonal unit cell,
\begin{align}
\frac{c_H}{a_H} = \sqrt{\frac{3}{2} \frac{1+2\cos\alpha_R}{1-\cos\alpha_R} }.
\end{align}
where $\alpha_R $ is the angle between any two vectors of  
a r-unit cell. 
$\alpha_R$ and $\alpha_C$ are related as follows:
\begin{align}
\cos \alpha_R = \frac{1}{2}\frac{1+3\cos\alpha_C}{1+\cos\alpha_C}.
\end{align}

We now derive expressions for the relevant temperature scales. 
From $Q_s$ of  Eq.~(\ref{eq:Qs}), we find that the stability
of the $r$ phase ends at the superheating temperature, 
\begin{align}
%T^* = T_0 + \frac{w_1 }{A_0}  \left( \frac{   \tilde{u} +   \tilde{v} }{ w_1 } \right)^2  + \left(\frac{2}{A_0}  \frac{e_a }{C_a} \right) P.
T^*(P) =  T^*(0)  + \left(\frac{2}{A_0}  \frac{e_a }{C_a} \right) P,
\end{align}
where $T^*(0) = T_0 + \frac{w_1 }{A_0}  \left( \frac{   \tilde{u} +   \tilde{v} }{ w_1 } \right)^2$
is the superheating temperature at ambient pressure. 
By equating the free energies $\tilde{G_Q}$ of the high and low temperature
phases, we find the transition temperature,
\begin{align}
\label{eq:Tc}
T_c(P) =  T_c(0) + \left(\frac{2}{A_0}  \frac{e_a }{C_a} \right) P,
\end{align}
where $T_c(0) = T_0 + \frac{3}{4} \frac{w_1 }{A_0} \left( \frac{   \tilde{u} +   \tilde{v} }{ w_1} \right)^2$ 
is the transition temperature at ambient pressure. 
In the next sections, we calculate several thermodynamic quantities of interest. 

\subsubsection{Coefficient of thermal expansion, entropy, latent heat, heat capacity, and barocaloric effect}

We begin with the volume change with temperature and the coefficient 
of thermal expansion (CTE).  
The temperature and pressure dependence of the volume $V$ is given by,~\cite{LevanyukFerroelectricity} 
\begin{align}
V(T) / V_0 =    \frac{\partial \tilde{G} }{\partial P}  =
\begin{cases}
     1 - \frac{P}{C_a},  & T>T_c \\
   1 - \frac{e_a}{C_a}Q_s^2 - \frac{P}{C_a},  & T<T_c. 
\end{cases} 
\end{align}
$V_0$ is a reference volume and
$ ( \partial G_0 / \partial P) =1$.
As expected, 
the relative change in volume in the r-phase is equal to 
the volumetric strain $\eta_1 + \eta_2 + \eta_2$.
The CTE is given as follows:
\begin{align}
	\kappa(T) &=  \frac{\partial^2  \tilde{G}  }{\partial T \partial P }, \nonumber \\
	 &=
\begin{cases}
   ( \partial V_0 /  \partial T ) = \kappa_0 ,  & T>T_c\\
  \kappa_0 +  \frac{1}{2} \frac{e_a }{C_a} \frac{A_0 /  w_1 }{\sqrt{     \left( \frac{   \tilde{u} +   \tilde{v} }{w_1} \right)^2 - \frac{\tilde{A}}{w_1}  } }  ,  & T<T_c.
\end{cases} %\nonumber
\end{align}

We calculate the entropy per unit volume from the free energy  of Eq. (\ref{eq:gibsRenormalized}),
\begin{align}
\label{eq:entropy}
S(T,P) = - \frac{\partial  \tilde{G}  }{\partial T} 
&= 
\begin{cases}
S_0, &  T > T_c, \\
S_0-\frac{A_0}{2}Q_s^2, & T < T_c.
\end{cases}
\end{align}

The latent heat per unit volume of the transition at $P=0$ is then given as follows,
\begin{align}
T_c \Delta S(T_c,0)=  T_c  \times \frac{ A_0}{2} Q_s^2(T_c,0),
 \end{align}
where $\Delta S(T_c,0)$ is the entropy jump at the $c-r$ transition.

We now calculate the heat capacity per unit volume,
\begin{align}
C_P &=  T  \frac{\partial S(T,P)}{\partial T}  \nonumber  \\
&= 
\begin{cases}
 C_P^0 , &  T > T_c, \\
 C_P^0 + T \frac{A_0}{4}  \frac{A_0 /  w_1 }{\sqrt{     \left( \frac{   \tilde{u} +   \tilde{v} }{ w_1 } \right)^2 - \frac{\tilde{A}}{w_1}  } }  , & T < T_c,
\end{cases}
\end{align}
where $C_P^0 = T \left( \partial S_0 / \partial T \right) $ is a reference heat capacity in the high temperature phase.

From Eq.~(\ref{eq:entropy}), we calculate the isotropic changes in entropy,
\begin{align}
\label{eq:BCE}
\Delta S(T,P) =
\begin{cases}
0, &  T > T_c, \\
-\frac{A_0}{2}\left(Q_s^2(T,P) -  Q_s^2(T,0) \right), & T < T_c.
\end{cases}
\end{align}

We conclude the presentation of our model here. In the next
section we apply it to several trifluorides.

\section{Results and Discussion}

\subsection{Fits and comparison to experiments}

\begin{figure}[htp!]
	\includegraphics[scale=0.325]{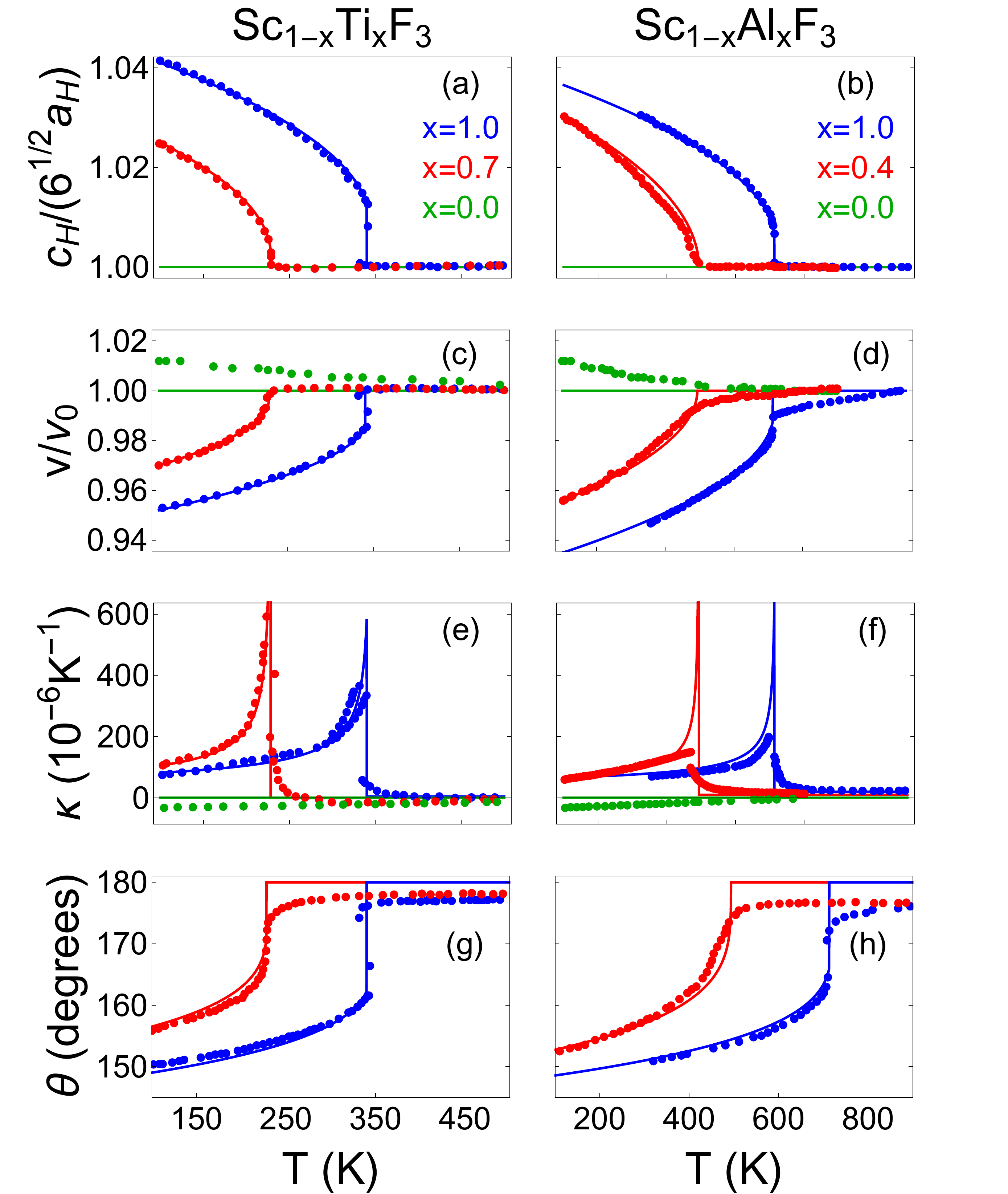}
	\caption{Comparison  between our model (solid line) and experiments (dots) for the temperature dependence of 
	(a,b) ratio hexagonal lattice constants, (c,d) unit cell volume, (e,f) CTE, and (g,h) metal-F-metal bond angle $\theta$
	in  Sc$_{1-x}$Ti$_x$F$_3$  and Sc$_{1-x}$Al$_x$F$_3$.
	Data taken from Refs.~[\onlinecite{Morelock2015a, Morelock2014a, Morelock2014b}].}
	\label{fig:comparison}
\end{figure}

We now discuss our fits to several trifluoride compounds.
For pure \tif\ and \alf ,
we fit our model to their observed $c_H/a_H$,  
M-F-M bond angle, volume expansion, CTE and 
latent heat of the transition,~\cite{Daniel1990a, Morelock2015a, Morelock2014a, Morelock2014b} 
see Fig.~\ref{fig:comparison}.
For Sc$_{1-x}$Al$_x$F$_3$ with $x<1$, we do a slightly different fit
since their latent heats are unknown: 
 we fix the ratio $dT_c / dP = 2 e_a / (A_0 C_a) $ to that of the pure compound \alf . 
This is justified by the observed linear  $T-P$ phase diagram 
of Sc$_{1-x}$Al$_x$F$_3$ with a slope that varies little with composition $x$.~\cite{Morelock2015a}
We do a similar fit for Sc$_{1-x}$Ti$_x$F$_3$.
The resulting parameters together with the calculated supercooling and superheating  temperatures 
are given in Table \ref{t:parameters}. 

Overall, we find that that there is good agreement between our model
and  experiments. 
The discrepancies between 
the observed and calculated M-F-M bond angles above $T_c$ shown
in  Figs.~\ref{fig:comparison} (g) and (h), 
are due to local lattice distortions from the average c-structure,~\cite{Chupas2004a, Chaudhuri2004a}
which we have not considered in our model. 
More importantly, 
the deviations from mean-field behavior are most noticeable 
in the volume expansions of  \scf\  and \alf ,
(see Figs.~\ref{fig:comparison} (c) and (d)),
which  correspond to 
the extreme cases of large and small
metal ion radius considered in this work~($r_\text{Sc}=0.745\,$\AA, $r_\text{Al}=0.535\,$\AA).
This suggests a trend with M-cation size. 
In \scf ,  NTE is the result of cooperative tilt fluctuations of the rigid ScF$_6$ octahedra
which reduce the average Sc-Sc distance while keeping the Sc-F distance fixed.  
Such fluctuations can only give rise to NTE; therefore the PTE in 
\alf\ must originate from
 non rigid modes such as Al-F bond stretching, which we
have not considered in our model. 
Sc$_{1-x}$Ti$_x$F$_3$ with $x=0.7$ has a mean B site radius 
in between these two extremes~($0.69\,$\AA)
and the deviations from our model and its observed volume expansion 
are tiny, which indicates that the fluctuations of both rigid 
	\begin{table}[hbp!]
		\begin{center} 
		\caption{Model parameters for Sc$_{1-x}$Al$_x$F$_3$ and Sc$_{1-x}$Ti$_x$F$_3$ and predicted supercooling ($T_0$) and superheating ($T^*$) temperatures at ambient pressure. Transition temperatures ($T_c$) taken from Refs.~[\onlinecite{Morelock2015a, Morelock2014a, Morelock2014b}]} 
		\label{t:parameters}
		\begin{tabular}{rrr|rr} \hline \hline
		                         &             \multicolumn{2}{c}{Sc$_{1-x}$Al$_x$F$_3$}         & \multicolumn{2}{c}{Sc$_{1-x}$Ti$_x$F$_3$}   \\ \hline
		                         &                $x=1.0$                & $x=0.4$               &      $x=1.0$                   & $x=0.7$ \\ \hline
			$\kappa_0\,$[$10^{-6}\,$ K$^{-1}$] & $ 10$   &  $10$           &  $ 0$                            &  $ 0$ \\
			$\tilde{u}+\tilde{v}\,$[meV  \AA$^{-7}$ ] & $-0.78$      & $0.05$      &  $-0.81$                    &   $-0.27$    \\
			$A_0\,$ [$10^{-3}\,$ meV  K$^{-1}$ \AA$^{-5}$] & $3.7$  & $5.8$ & $2.9$ &  $3.1$     \\
            $w\,$[ meV \AA$^{-9}$] & $12$  &                  $48$                       &  $13$                          &  $ 17 $\\
           	$ e_a  / C_a\,$[ \AA$^{-2}$] & $0.13\,$              &  $0.20$     &  $0.17$                                & $0.17$\\
             $ e_r  / C_r\,$[\AA$^{-2}$] & $0.14$       &          $0.27$             &   $0.28$                            & $ 0.30$ \\ \hline 
             $T_0\,$[K]                          & $ 703 $     & $493$                       & $327$                               &  $227$ \\ 
             $T^* \,$[K]                          & $718$       &  $493$                       & $344$                               & $228$ \\  \hline
             $T_c \, $[K]                          & $713$        &  $493$                     & $ 340 $                             & $228$  \\ \hline \hline
		\end{tabular}
		\end{center}
	\end{table}
and non-rigid modes are unimportant.
A picture therefore emerges in which rigid octahedra fluctuations dominate the TE for
large metal ions and decrease with their size, while non rigid vibrational modes
dominate the TE for small metal ions and decrease with increasing radius.

\begin{figure}[htp]
		\includegraphics[scale=0.325]{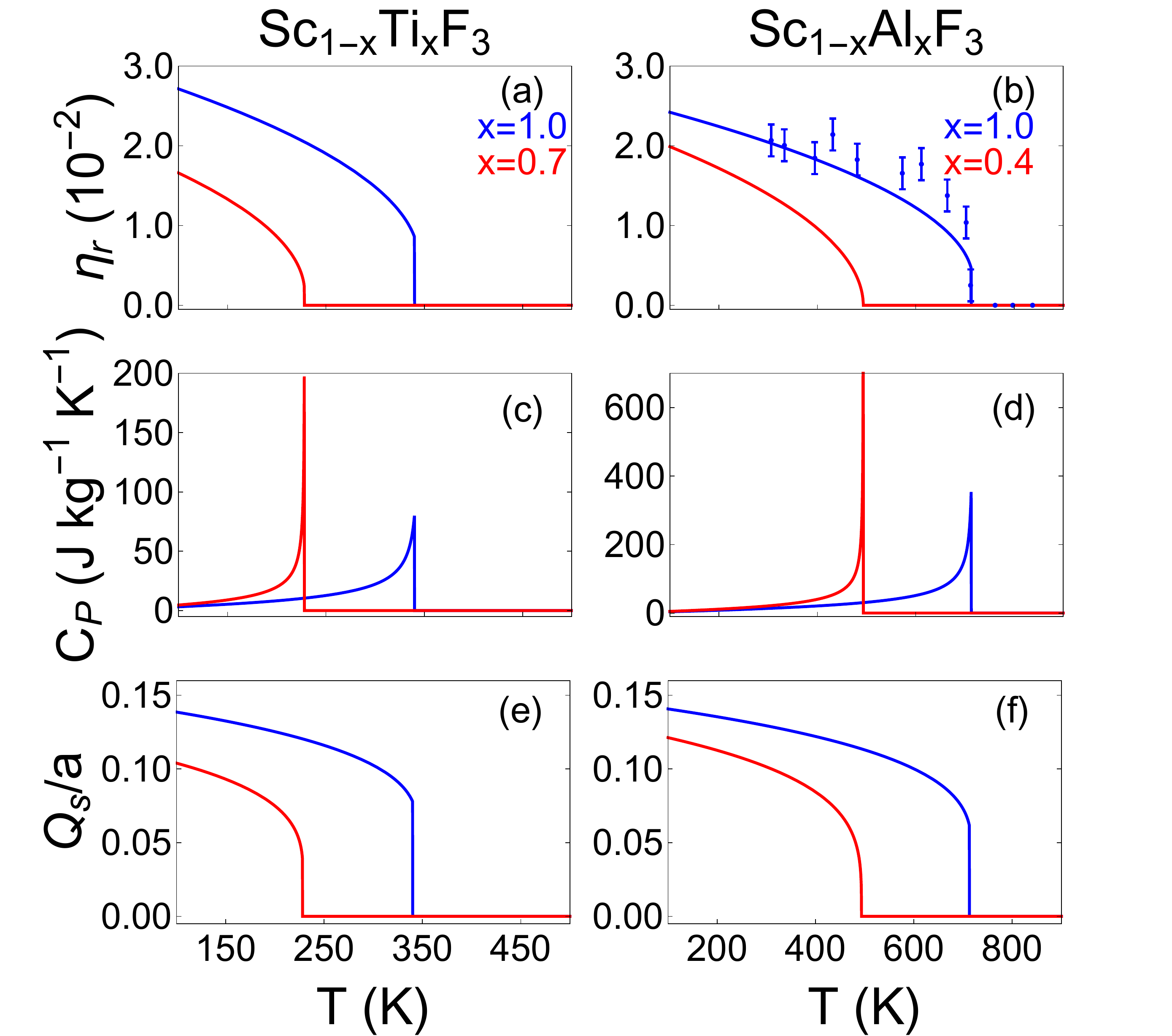}
    \caption{(a,b) Sponteanous shear strains, (c,d) excess specific heat, and (e,f) 
	order parameter predicted from the fits obtained from 
	Fig.~\ref{fig:comparison} for Sc$_{1-x}$Ti$_x$F$_3$  and Sc$_{1-x}$Al$_x$F$_3$.  
	Measured spontaneous strains (dots) taken from Ref.~[\onlinecite{Kennedy2002a}].}
	\label{fig:calculated}
\end{figure}

Fig.~\ref{fig:calculated} shows the predicted 
 spontaneous shear strains, order parameter, and specific heats
for Sc$_{1-x}$Al$_x$F$_3$  and Sc$_{1-x}$Ti$_x$F$_3$ using 
our parametrization. 
Our prediction for the shear strains  in \alf\ compares well with experiments.~\cite{Kennedy2002a}

\begin{figure}[htp]
	\centering
	\includegraphics[scale=0.4]{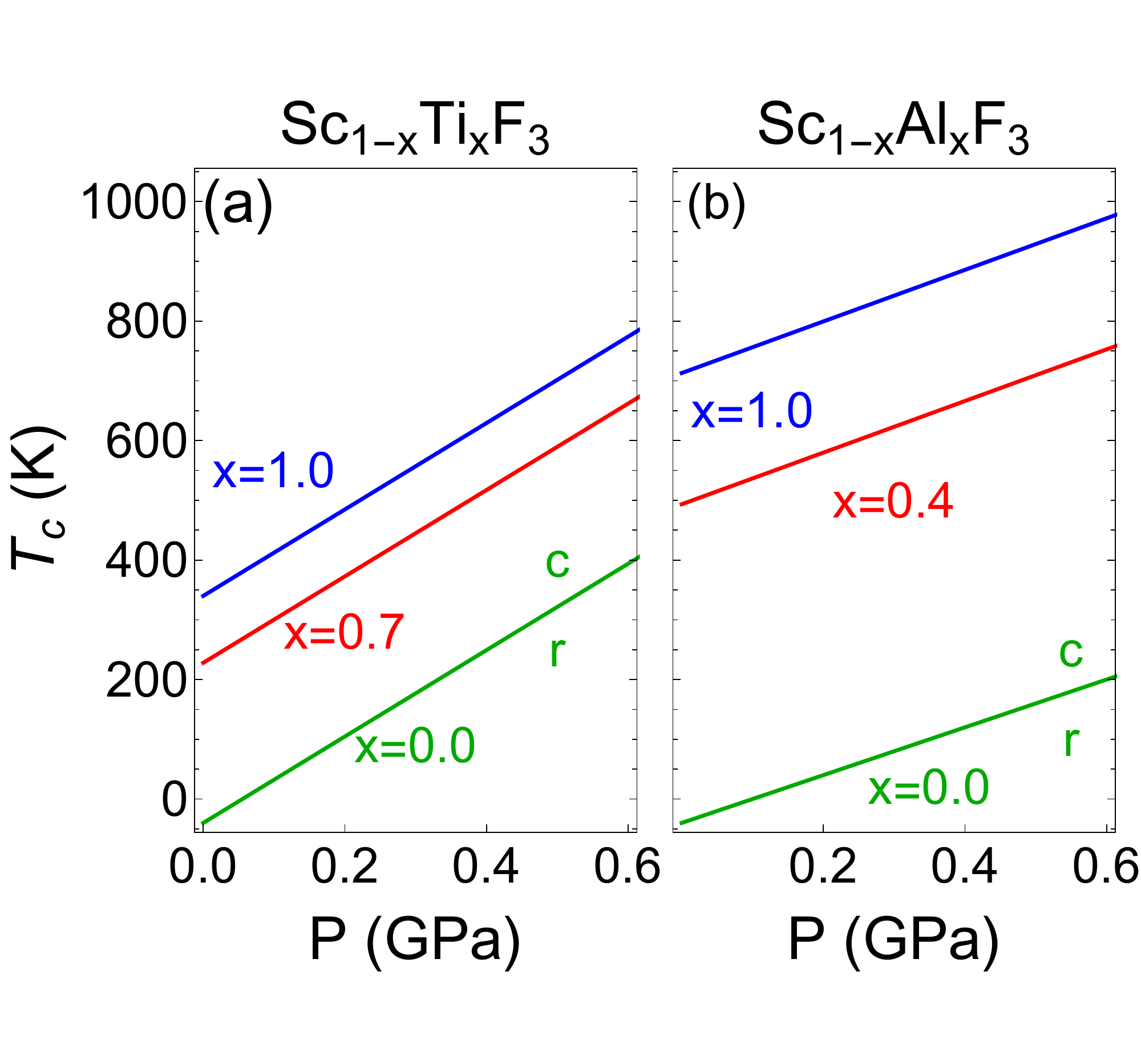}
	\caption{Calculated $T-P$ phase diagrams for (a)~Sc$_{1-x}$Ti$_x$F$_3$ and (b)~Sc$_{1-x}$Al$_x$F$_3$.}
	\label{fig:PTPhaseDiagrams}
\end{figure}

Figures~\ref{fig:PTPhaseDiagrams}~(a) and~(b) show the $T-P$ phase diagrams of
Sc$_{1-x}$Ti$_x$F$_3$ and Sc$_{1-x}$Al$_x$F$_3$ calculated from Eq.~(\ref{eq:Tc}).
The linear dependence between $T$ and $P$ of our model agrees with 
the observed phase diagram for Sc$_{1-x}$Al$_x$F$_3$.~\cite{Greve2010a, Morelock2015a}
Using Eq.~(\ref{eq:Tc}) and the parameters from Table~\ref{t:parameters}, we 
find that $ dT_c / dP \simeq 0.4 \times 10^{3}\,$ K GPa$^{-1}$
which is in excellent agreement with experiments.~\cite{Greve2010a, Morelock2015a}
For Sc$_{1-x}$Ti$_x$F$_3$, we predict that $  dT_c / dP \simeq  0.7 \times 10^{3}\,$ K GPa$^{-1}$.
There are no reports on the calculated or measured  $T-P$ phase diagram for Sc$_{1-x}$Ti$_x$F$_3$  in the literature.

In Appendix A, we derive expressions for the temperature dependence of the soft mode frequencies.
In the parent $c$ phase, we find the usual mean field behavior 
for the $R_4^+$ mode, $\omega_{R_4^+} \propto \sqrt{A_0 (T-T_0)}$, which is  
in agreement with IXS.~\cite{Handunkanda2015a}
In addition, our values for $A_0$  given in Table~\ref{t:parameters} are 
about what is expected from the observed soft mode (
  $A_0 \simeq  3 \times 10^{-3}\,$  meV  K$^{-1}$ \AA$^{-5}$) for \scf .~\cite{Handunkanda2015a}
The temperature dependence of the A$_{1g}$ and E$_g$ phonon frequencies in the r phase 
has been measured  by Raman scattering experiments for \alf ,~\cite{Daniel1990a}
however, we cannot compare to our model as the shear moduli $C_t$ and $C_r$ are unknown. 
\begin{table*}[!]
  \caption{Transition temperatures ($T_c$), isothermal entropy changes ($\Delta S$), isothermal heats ($\mathcal{Q}=T_c \Delta S $), 
  pressure changes ($\Delta P$), caloric strengths ($\Delta S / \Delta P$), refrigerant capacity~(RC), and $T-P$ slope~($dT_c/dP$) of giant barocaloric materials.}
  \begin{tabular}{lcccccccc} \hline
     Compound & $T_c\,$[K] &   $\Delta S\,$[JK$^{-1}$kg$^{-1}$] & $\mathcal{Q}\,$[kJkg$^{-1}$]  & $\Delta P\,$[GPa] & $ \frac{\Delta S}{ \Delta P}\,$[JK$^{-1}$kg$^{-1}$GPa$^{-1}$] &RC [Jkg$^{-1}$] & $\frac{d T_c}{ dP}\,$[KGPa$^{-1}$] &  Ref.  \\ \hline \hline
      Ni$_{49.26}$Mn$_{36.08}$In$_{14.66}$  & $293$ & $24$ & $7.0$ & $0.26$ & $92$ & $120$ &$18$ &\onlinecite{Manosa2010a}  \\ 
      LaFe$_{11.33}$Co$_{0.47}$Si$_{1.2}$ & $237$ &  $8.7$ &$2.0$ & $0.20$ & $43.5$ & $81$ & $73$  & \onlinecite{Manosa2011a} \\
(NH$_4$)$_3$MoO$_3$F$_3$ & $297$ & $55 $ & $16.3$ & $0.5$ & $ 110 $ & $5200$ & $202$ &\onlinecite{Flerov2011a}  \\  
       Gd$_5$Si$_2$Ge$_2$  & $ 270 $ & $11$& $2.9$ & $0.20$ & $55$ &$180$ & $32$ & \onlinecite{Suheyla2012a} \\
     Fe$_{49}$Rh$_{51}$ & $308$ & $12.5$ & $3.8$ & $0.11$ & $ 114 $ & $105$ & $54$ & \onlinecite{Stern2014a} \\
           Mn$_3$GaN & $285$  & $21.6$ & $6.2$  & $0.09$  & $240$ &$125$ & $65$ & \onlinecite{Matsunami2015a} \\
           (NH$_4$)$_2$SO$_4$           &   $  219 $  & $ 60 $  & $13.2$ & $ 0.10 $   &  $600$& $276$  & $45$ & \onlinecite{Lloveras2015a} \\
         BaTiO$_3$          &   $ 400 $  & $1.6 $  &  $ 0.64 $ & $ 0.10 $   &  $16$ & $10$  &\hspace{-0.25cm}$-58$ & \onlinecite{Stern-Taulats2016a} \\      
         $[$TPrA] [Mn(dca)$_3$]    & $330$ &  $35.1$ & $11.6$ & $0.00689$ & $ 5094 $& $62$ & $231$ & \onlinecite{BermudezGarcia2017a} \\ 
  Sc$_{1-x}$Ti$_x$F$_3\,(x=0.85)$         &   $ 283 $  & $12$ & $ 3.4 $  & $ 0.10 $   &  $ 120$ &$ 406$  & $ 723$ & This work \\ \hline
    \end{tabular}
   \label{t:barocalorics}
\end{table*}

\subsection{Barocaloric effect}

\begin{figure}[htp!]
	\centering
	\includegraphics[scale=0.55]{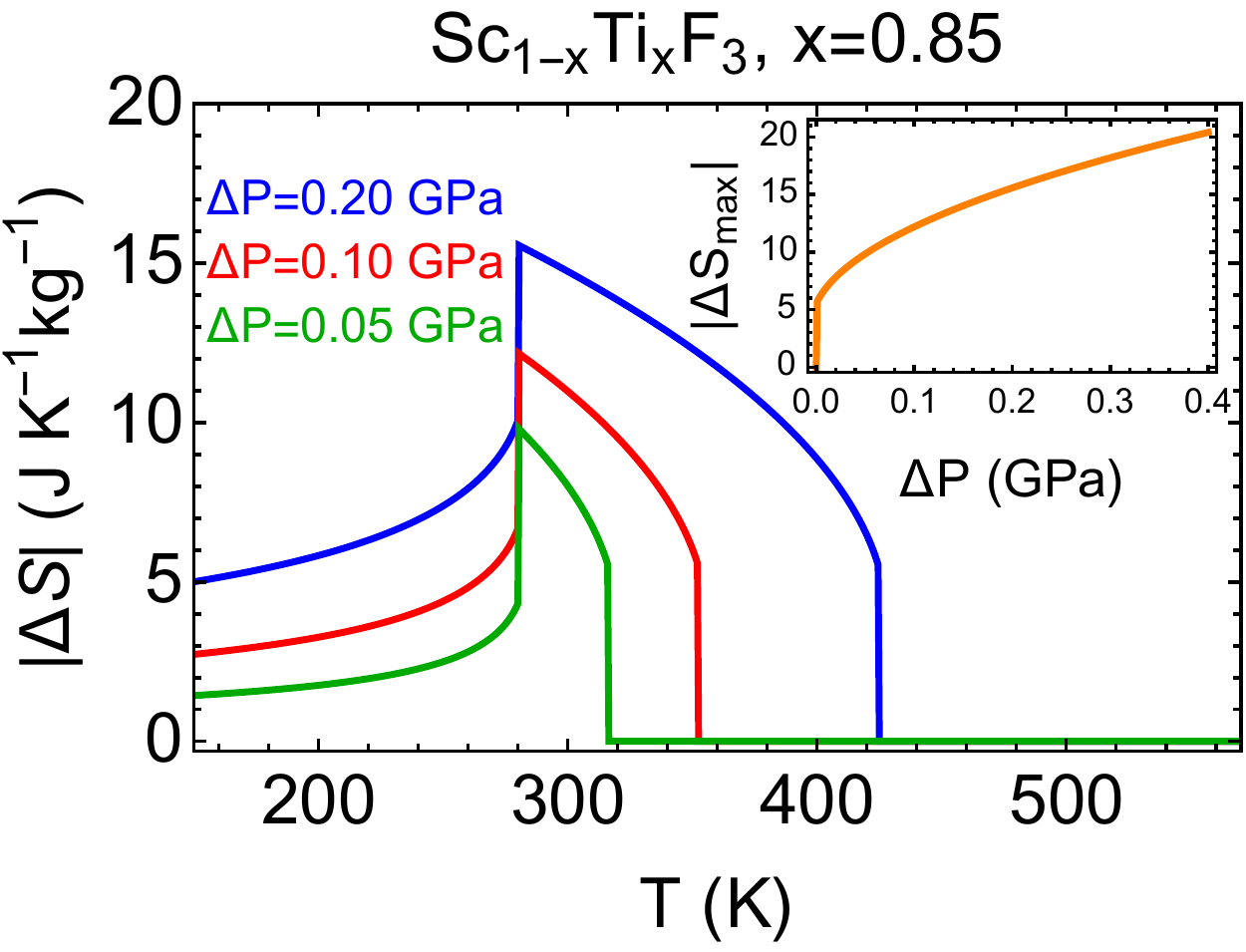}
	\caption{Calculated isothermal changes in entropy for  Sc$_{1-x}$Ti$_{x}$F$_3$, $x=0.85$. 
	Inset: Pressure dependence of the maximum changes in entropy.}
	\label{fig:BCE}
\end{figure}

Figure~\ref{fig:BCE} shows the predicted BCE for Sc$_{1-x}$Ti$_x$F$_3\,(x=0.85)$ calculated
from Eq.~(\ref{eq:BCE}).
We have chosen this composition as its $c-r$ transition occurs near RT ($T_c=283\,$K) and 
exhibits a strong first order character.~\cite{Morelock2014a} 
As expected from the large CTEs, the resulting isothermal changes in entropy are comparable
to those exhibiting giant effects,~\cite{Lloveras2015a} as it is shown in Table \ref{t:barocalorics}.
 Moreover, the effect extends over a temperature range of about  $140\,$K
for pressure changes of $0.2\,$GPa which includes RT.
The wide temperature range 
is a consequence of the large predicted value of $dT_c / dP$($=723\,$KGPa$^{-1}$), which exceeds those of 
typical barocaloric compounds, see Table~\ref{t:barocalorics}.
The inset in Fig.~\ref{fig:BCE} shows 
the expected monotonic growth of maximum entropy changes at $T_c$, $\Delta S_{max}$,
with changes in pressure. 

\subsection{Comparison to \reo\ }

\begin{figure}[htp!]
	\centering
	\includegraphics[scale=0.6]{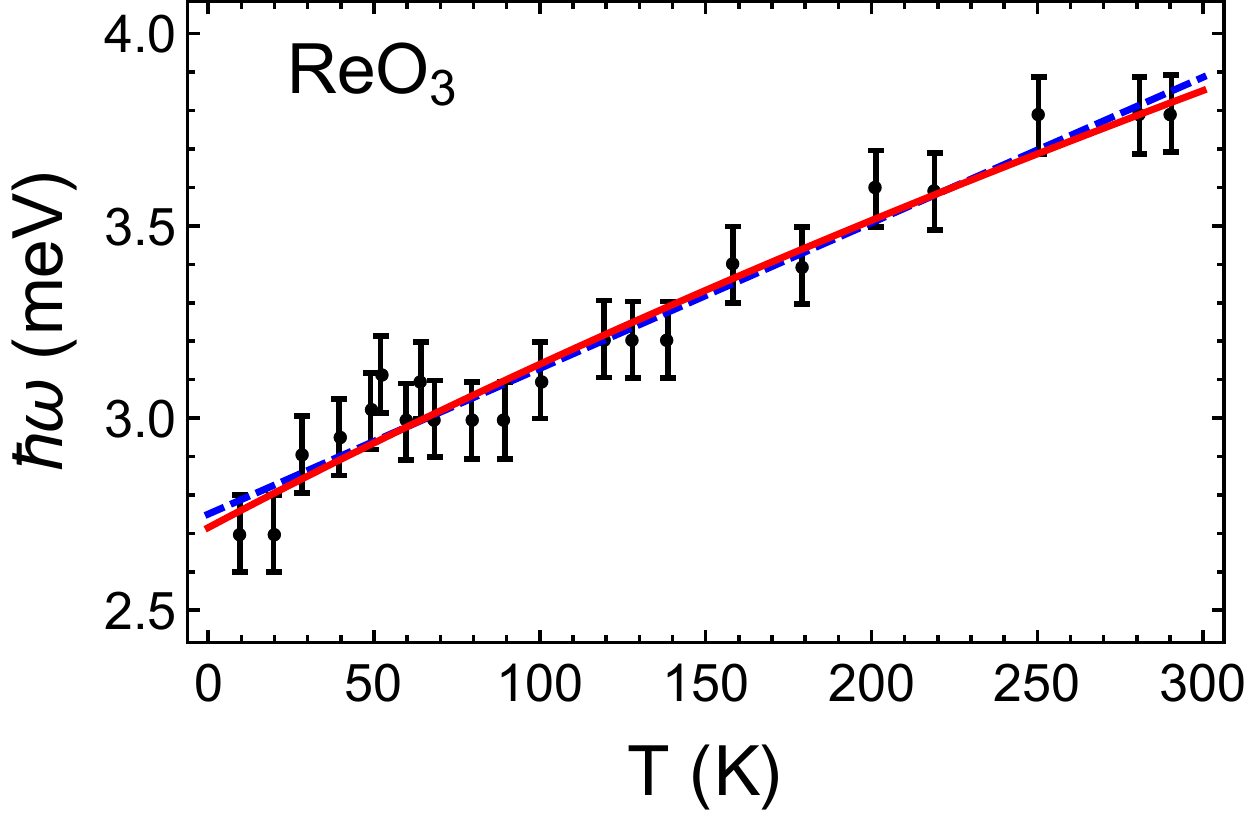}
	\caption{Temperature dependence of the M$_3^+$ phonon mode in \reo . 
	Dashed-blue and solid-red lines correspond to purely linear and classical mean-field behavior, respectively.
	Data taken from Ref.~[\onlinecite{Chatterji2009b}] }
	\label{fig:ReO3}
\end{figure}

It is interesting to compare \scf\ with the isostructural compound
\reo .  At ambient pressure, \reo\ exhibits a perovskite $c$ lattice structure  from 
the lowest observed temperature up to its melting point
despite its empty A site and therefore low tolerance factor.~\cite{Tsuda2000a}   
The stability of the $c$ phase is a consequence of its metallicity:
the Fermi pressure of delocalized Re $5d$ electrons that occupy 
the $\pi^*$ conduction band keep the ReO$_6$ octahedra from tilting.~\cite{Stachiotti1997a} 
On the other hand,  such states are empty
in the wide-gap insulator ScF$_3$.~\cite{Bocharov2016a, *Hamed2015a}  
Its lattice structure remains cubic at all temperatures  due to its purely ionic Madelung energy.~\cite{Chen2004a}
Both compounds exhibit incipient lattice instabilities in their c-phases.
In \scf, softening of the entire M-R phonon branch (with lowest point at R) 
has been observed by IXS~\cite{Handunkanda2015a}  from $300$ to $8\,$K at ambient pressure and also 
found by a first-principles calculation.~\cite{Roekeghem2016a}
The temperature dependence of the phonon energies is well-described by mean-field theory, as discussed above. 
Condensation of the R$_4^+$ mode and its associated $c-r$ transition 
can be induced by modest hydrostatic compression~($\sim 0.6\,$GPa at RT)~\cite{Greve2010a, Aleksandrov2009a, Aleksandrov2002a} 
or cation substitution.~\cite{Morelock2013a, Morelock2014a, Morelock2015a}
It is unknown whether the transition is of first or second order. 
In \reo , inelastic neutron scattering experiments~\cite{Chatterji2009b} ~(INS) have observed 
softening  from $280$ to $2\,$K without condensation of the M$_3^+$ phonon 
mode, which consists of in-phase rotations of rigid ReO$_6$ octahedra 
along  a [$100$] axis that passes through the metal cation. 
The supposedly observed linear temperature dependence of the mode energy is unusual
as it is shown by the blue dashed line in Fig.~\ref{fig:ReO3}.
However, we make the observation that the linear fit  is hardly distinguishable from the standard
mean-field behavior, $\omega_{M_3^+} \propto \sqrt{A_0 \left(T-T_0\right)}$, with physically 
reasonable parameters~($A_0 \simeq 2.5 \times 10^{-3}\,$ meV  K$^{-1}$ \AA$^{-5}$, and $T_0 \simeq -296\,$K),
see solid red line in Fig.~\ref{fig:ReO3}.
The mode can be condensed upon application of moderate
pressures~($\sim 0.5\,$GPa at RT)  and a $P-T$ phase diagram  has been 
established by neutron diffraction experiments~\cite{Chatterji2006a}
in which the high-pressure phase has $c$ symmetry ($Im3$)
and the transition line is of second-order.~\footnote{A recent first-principles calculation~\cite{Muthu2015a} has suggested that the c-c transition is weakly first-order.}
Additional pressure-induced transitions have been reported in \scf~\cite{Aleksandrov2009a, Aleksandrov2002a}  
and \reo\ ~\cite{Suzuki2002a, *Jorgensen2000a} at higher pressures.

Both \scf ~\cite{Greve2010a} and \reo ~\cite{Rodriguez2009a, Chatterji2009a, Chatterji2008a} exhibit negative 
TE over a wide temperature range
with a common origin: 
large quartic anharmonicities of their corresponding soft R$_4^+$ and M$_3^+$ modes
consisting of rigid antiphase rotations of the ScF$_6$ and ReO$_6$ octahedra, respectively.~\cite{Li2011a, Chatterji2009b}
The size of the effect, however, is an order 
of magnitude larger in \scf\ than in \reo .
This can be understood from the distinct nature of their metal-nonmetal bonds.
In \scf , there is little overlap between the 
charge densities that form the ionic bond between
Sc$^{3+}$ and F$^{-}$,~\cite{Liu2015a} which favors large buckling fluctuations in the Sc-F-Sc chains.
In \reo , the buckling fluctuations of the Re-O-Re bonds are reduced by 
 the stiffer covalent bond formed by hybridized
Re $5d$ and O $2p$ electrons.~\cite{Cora1997a, Stachiotti1997a}

\subsection{Comparison to \wo }

Another interesting compound to compare with is tungsten
bronze \wo .
Like \scf , \wo\ is an insulator but its high temperature phase
is tetragonal~($t$) and it goes through several structural transitions upon cooling.~\cite{Tsuda2000a}
Its hypothetical $c$ structure has an unstable M$_3^{-}$ mode
consisting of oposite displacements of the
cations and anions from unit cell to unit cell  
along the [110] directions,~\cite{Hamdi2016a} which generate off-center displacements
of W$^{6+}$ towards one of its nearest O$^{2-}$ with concomintant
increase in their covalency.~\cite{Cora1996a}  
Condensation of this mode leads to a t-structure of 
highly distorted WO$_6$ octahedra.\cite{Kehl1952a}   
The energy gain due to the increase in covalency between 
W$^{6+}$ and O$^{2-}$ favors the $t$ phase over 
the ionic c-structure.~\cite{Cora1996a} 
The $c$ phase can be stabilized in \wo\ by introducing electrons: 
when doped with Na,  $3s$ electrons begin to occupy the conduction band
and, for sufficiently large  concentrations, their Fermi pressure stabilizes 
the $c$ phase.~\cite{Cora1997a} Such a $c$ phase displays PTE
and mean-field softening with temperature of its M$_3^-$ phonon mode  down to 
about $416\,$K where a structural transition to a $t$ phase occurs.~\cite{Sato1982a} 

\section{Conclusions}

We have presented a Landau theory for trifluoride and have used it to calculate and predict 
the temperature and pressure dependence
of several thermodynamic quantities.
We have compared our results to existing experimental 
data on trifluorides and have quantified the deviations from
mean-field theory. We have found that the fluctuations
of their rigid MF$_6$ octahedra 
tend to  increase with the metal cation size. 
 We have used our model to 
predict a giant BCE in Sc$_{1-x}$Ti$_x$F$_3$~($x=0.85$) 
of up to $15\,$JK$^{-1}$kg$^{-1}$ for a pressure change of 
$0.2\,$GPa. 
This effect extends over a temperature range of over $140\,$K, which
includes RT.
Our results suggest that open lattice frameworks such as the trifluorides could be a promising 
platform to search for giant barocaloric effects. 

\section{Acknowledgments}
GGGV thanks Xavier Moya and Esteban Aveda\~{n}o for useful discussions
and Jason Hancock and Sahan Handunkanda for carefully reading the manuscript. 
Work at the University of Costa Rica is supported by the Vice-rectory for Research
under the project no. 816-B5-220, and work at Argonne National Laboratory is supported by the U.S. Department of Energy, 
Office of Basic Energy Sciences,  Material Sciences and Engineering Division under contract no. DE-AC02-06CH11357.
GGGV and RTB acknowledge partial financial support from
the Office of International Affairs at the University of Costa Rica.

\appendix

\section{Soft Mode Frequencies}

The soft mode frequencies are computed from
the free energy~(\ref{eq:gibs}) with the $\eta$'s constant,~\cite{Slonczewski1970a} 
\begin{align}
\label{eq:modesHF}
  \varrho \left( \omega_i \right)^2 \delta_{ij} = \frac{\partial^2 (G_Q+G_{\eta}) }{\partial \hat{Q}_i \partial \hat{Q}_j},~~(i,j=1,2,3)
\end{align}
where $\hat{Q}_i$ are principal-axis coordinates of the soft mode and $\varrho=2 m_F/a^3$
is the mass density of fluorine atoms participating in each mode, where
$m_F$ is the mass of the fluorine atom. 
The soft mode frequencies given in Eq.~(\ref{eq:modesHF}) must be evaluated at the equilibrium points
given in Eqs.~(\ref{eq:Qs}) and~(\ref{eq:etas}).  
The free energy $G$ appearing in Eq.~(\ref{eq:modesHF}) is that of
Eq.~(\ref{eq:gibs}) rather than that of Eq.~(\ref{eq:gibsRenormalized})  
because the frequency of the  acoustic modes associated with 
uniform strains vanishes in the long-wavelength limit.~\cite{Slonczewski1970a}

In the $c$ phase, the frequency of the $R_4^+$ mode is threefold degenerate since all 
strains vanish,
\begin{align}
\label{eq:R5}
\varrho \,  \omega^2_{R_4^+}  = \tilde{A}.
\end{align}
In the $r$ phase, the mode splits into the $E_g$ doublet and the $A_{1g}$ singlet, 
\begin{subequations}
\label{eq:softmodes}
\begin{align}
\label{eq:Eg}
\varrho \, \omega^2_{E_g}&=  \tilde{A}+\left( 2 \tilde{u} +8 \frac{e_a^2}{C_a}+ 6 \frac{ e_t^2 }{  C_t } + \frac{1}{3} \frac{ e_r^2 }{ C_r } +  w_1 \, Q_s^2  \right)  Q_s^2, \\
\label{eq:A1g}
\varrho \, \omega^2_{A_{1g}} &=  \tilde{A}+\left( 6 \left[ \tilde{u} +   \tilde{v} \right] + 28 \frac{ e_a^2 }{  C_a } + \frac{4}{3} \frac{ e_r^2 }{ C_r }  + 5  w_1 \, Q_s^2  \right) Q_s^2,  
\end{align}
\end{subequations}
where  $Q_s$ is given in Eq. (\ref{eq:Qs}).

\section{Sixth-order $c$ anisotropy}

In this Appendix, we discuss the effects of sixth-order anisotropies in some of our
previous results. We will show that such anisotropies allow us to describe 
a possible phase competition between pressure-induced phases.
So far,the  evidence for phase competition
has been experimentally reported in \scf\,~\cite{Aleksandrov2002a, Aleksandrov2009a}  
where near about $3.0\,$GPa, the $r$ phase destabilizes and a structural transition to an 
orthorhombic (o) phase occurs.
In addition, a MD simulation of \alf\ has 
found a metastable o-phase in the free energy at ambient pressure and below 
the $c-r$ transition temperature.~\cite{Chaudhuri2004a}
No $r-o$ transition has been observed 
in \tif , FeF$_3$, and CrF$_3$.~\cite{Sowa1998a, Jorgensen2004a} 

It is well known that the free energy (\ref{eq:gibsRenormalized}) does not support a stable $o$ phase.~\cite{Cowley1980a} 
To include it, we must add sixth-order $c$ anisotropies,
\begin{align}
\label{eq:6thOrderAn}
 \frac{3w_2}{4} \left( Q_1^2 (Q_2^4 + Q_3^4) +  Q_2^2 ( Q_1^4 + Q_3^4)  \right. \nonumber \\
  \left. + Q_3^2 (Q_1^4 + Q_2^4)  \right) \\
+ \frac{9w_3}{2} Q_1^2 Q_2^2 Q_3^2. \nonumber
\end{align}
where $w_2$ and $w_3$ are parameters independent of temperature and pressure.
We consider the following order parameters for the t-, o- and  r-phases,
\begin{align*}
t&: (Q_1, Q_2, Q_3) = Q_s \left( 0, 0 , 1 \right), \\
o&: (Q_1, Q_2, Q_3) = \frac{Q_s}{\sqrt{2}}  \left(1 ,1 , 0 \right), \\
r&: (Q_1, Q_2, Q_3) = \frac{Q_s}{\sqrt{3}}  \left(1 ,1 , 1 \right). 
\end{align*}

The contribution from the anisotropic terms of Eq.~(\ref{eq:6thOrderAn}) to the free energy is,
\begin{subequations}
\begin{align}
\tilde{G}_{AN} (\textrm{t}) &=  0, \\
\tilde{G}_{AN} (\textrm{o}) &= \left( \frac{3}{8} \tilde{v} + \frac{3}{16}  w_2  Q_s^2 \right) Q_s^4, \\
\tilde{G}_{AN} (\textrm{r}) &=  \left( \frac{1}{2} \tilde{v} + \frac{1}{6} \left(  w_2 + w_3 \right) Q_s^2 \right) Q_s^4.
\end{align}
\end{subequations}

\begin{figure}[htp!]
	\centering
	\includegraphics[scale=0.425]{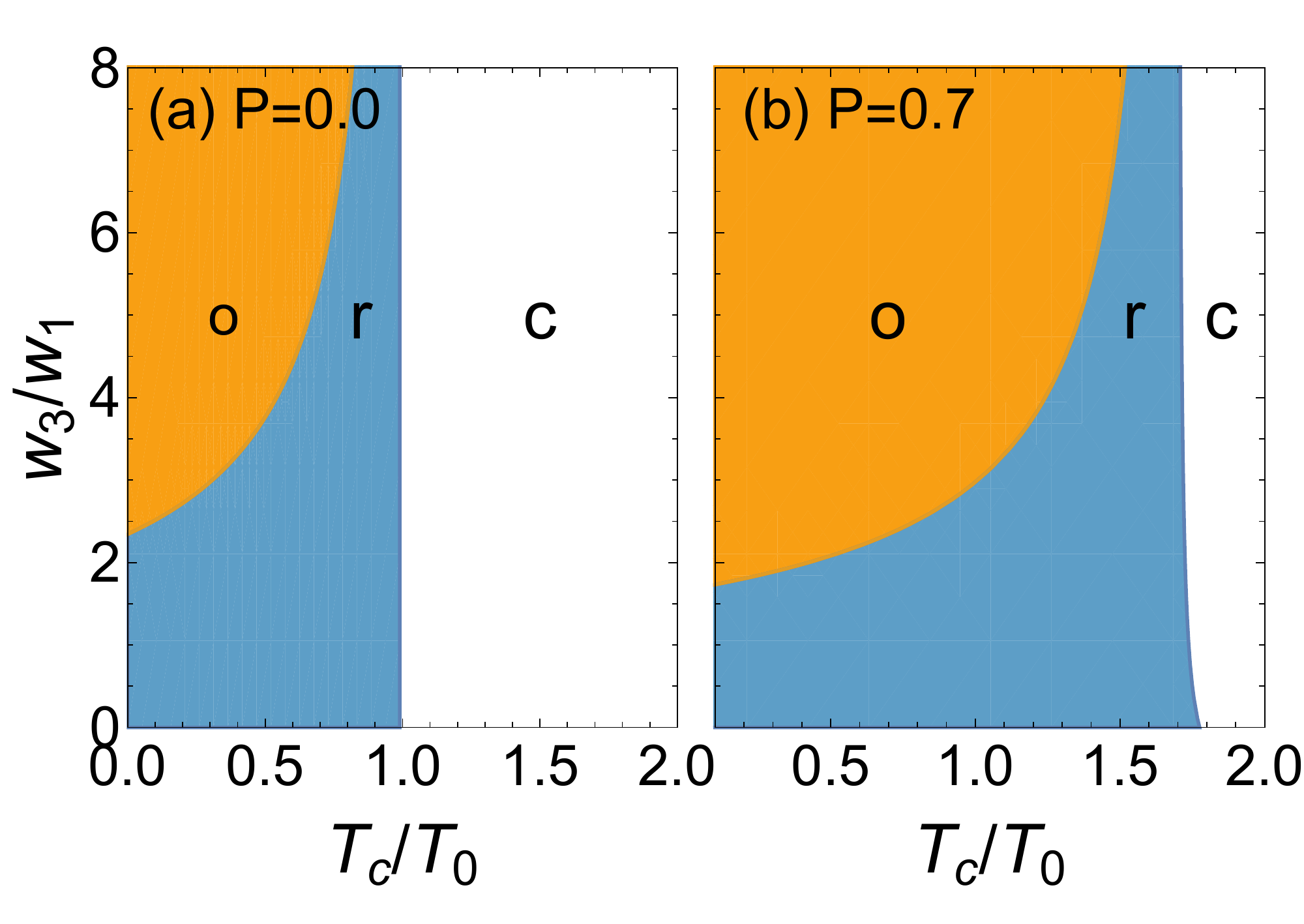}
	\caption{Schematic phase diagrams of the trifluoride (a) at ambient pressure and (b) with applied hydrostatic compression. 
				All transition lines are of first-order. Here, $A_0 T_0 / (w_1 a^4)= 1.0 \times 10^{-3},\, \tilde{u}/(w_1 a^2)= 6.0\times 10^{-2}$, 
				and  $\tilde{v}/(w_1 a^2) = -7.0 \times 10^{-2}$. $P$ is in units of $C_a A_0 T_0 / (2 e_a)$.}
	\label{fig:phasediagrams}
\end{figure}

\begin{figure}[htp!]
\centering
\includegraphics[scale=0.5]{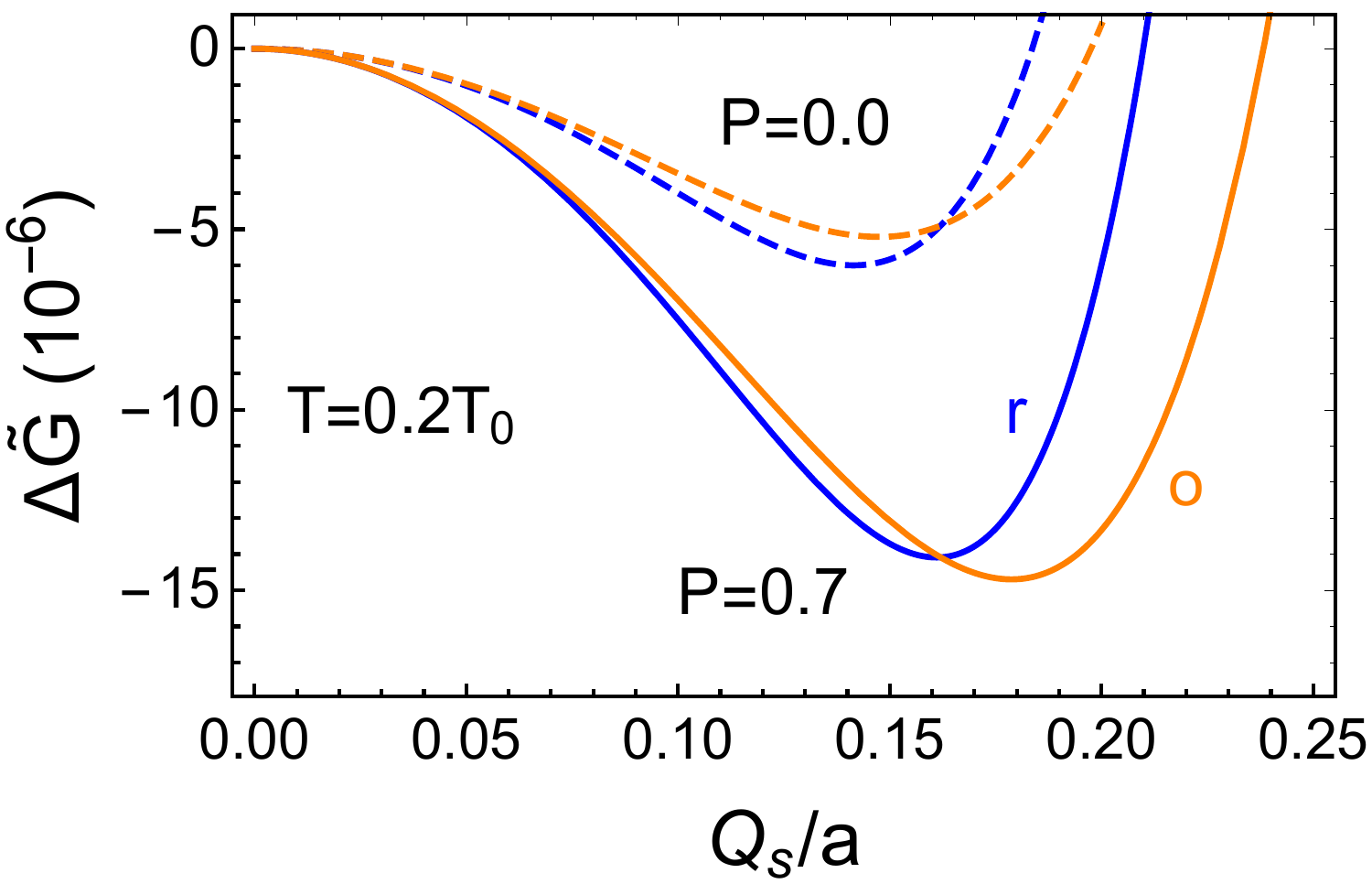}
\caption{Free energy densities for $w_3 / w_1 = 2.0$. $\Delta \tilde{G}$ and $P$ are in units of $w_1 a^6$ and $C_a A_0 T_0 / (2 e_a)$, respectively.}
\label{fig:freeenergy}
\end{figure}

For simplicity, we take $w_2=0$. 
Then, for $\tilde{v}<0$ and $  w_1 + w_3  > 0 $, 
we find that the  $r$ phase is the global minimum for small spontaneous distortions~($Q_s^2  <  - (3/4)(\tilde{v}/w_3) $) 
while the $o$ phase is a local minimum; and viceversa for large distortions~[$Q_s^2 >  - (3/4)(\tilde{v}/w_3) $].
For  $\tilde{v}<0$ and $ w_1 +  w_3  \leq 0 $, the energy  has unphysical divergences implying that
higher-order terms must be taken into account. 
The $t$ phase is always metastable.

The order parameter of the $o$ phase with $w_2=0$ is given as follows,
\begin{align*}
Q_s(T,P) =\pm \left\{ \sqrt{\left( \frac{\tilde{u} + 3 \tilde{v} / 4 }{w_1   } \right)^2 - \frac{\tilde{A}}{w_1 } } - \left( \frac{\tilde{u} + 3\tilde{v}/4 }{w_1 } \right) \right\}^{1/2},
\end{align*}
where $\tilde{A}$ is given by Eq.~(\ref{eq:ARenormalized}). The order parameter of the r-phase
with sixth-order anisotropies is obtained by replacing $w_1 \to w_1 + w_3$ in Eq.~(\ref{eq:Qs}).

Figure \ref{fig:phasediagrams}\,(a) shows a generic $w_3-T$ phase diagram at ambient pressure calculated by 
comparing the free energies of the $c$, $r$ and $o$ minima. For small anisotropies~($w_3 / w_1 \lesssim 2.2$), we find 
there is only a $c-r$ phase transition; while for large anisotropies~($w_3 / w_1 \gtrsim 2.2$), an
additional $r-o$ phase change occurs at low temperatures. 
Figure \ref{fig:phasediagrams}\,(b) shows the $w_3-T$ phase diagram at an applied pressure. 
As expected, pressure favors ordering:
as the $c-r$ and $r-o$ transitions are pushed to higher temperatures. 
The corresponding free energy changes $\Delta \tilde{G} = \tilde{G} - G_0 +P^2 / (2C_a)$ of the 
$r$ and $o$ phases for $w_3 / w_1 = 2.0$ are shown in 
Fig.~ \ref{fig:freeenergy}. The free energy of the metastable $t$ phase is not shown for clarity. 
At ambient pressure, the $r$ phase is a global minimum while the $o$ phase is metastable, 
which is in agreement with MD simulations.~\cite{Chaudhuri2004a} With large enough applied pressures,
the situation reverses and the r-phase becomes a local minimum while  the $o$ phase is the ground state. 
Metal trifluorides must then lie in the region of low anisotropy~($w_3 / w_1 < 2.2$), as 
no transition to an $o$ phase has been observed at ambient pressure.

%\bibliography{references}

%

\end{document}